\documentclass[dvipdfmx,aps,prd,preprintnumbers,superscriptaddress,nofootinbib,showpacs,twocolumn]{revtex4}%
\usepackage[dvipdfmx]{graphicx}
\usepackage{bm,latexsym,amsmath,amssymb,amsfonts,mathrsfs}
\usepackage{color}
\input{colordvi.tex}
\usepackage[dvipdfmx]{hyperref}
\usepackage{url}
\hypersetup{
    colorlinks=true,
    citecolor=cyan,
}
\newcommand*{\D}{{\rm d}}
\newcommand*{\mpl}{M_{\rm Pl}}
\begin{document}

\title{Ultra slow-roll G-inflation}

\author{Shin'ichi~Hirano}
\email[Email: ]{s.hirano"at"rikkyo.ac.jp}
\affiliation{Department of Physics, Rikkyo University, Toshima, Tokyo 171-8501, Japan}
\author{Tsutomu~Kobayashi}
\email[Email: ]{tsutomu"at"rikkyo.ac.jp}
\affiliation{Department of Physics, Rikkyo University, Toshima, Tokyo 171-8501, Japan}
\author{Shuichiro~Yokoyama}
\email[Email: ]{shuichiro"at"rikkyo.ac.jp}
\affiliation{Department of Physics, Rikkyo University, Toshima, Tokyo 171-8501, Japan}

\begin{abstract}
The conventional slow-roll approximation is broken in
the so-called ``ultra slow-roll'' models of inflation,
for which the inflaton potential is exactly (or extremely) flat.
The interesting nature of (canonical) ultra slow-roll inflation is that
the curvature perturbation grows on superhorizon scales, but has a scale-invariant power spectrum.
We study the ultra slow-roll inflationary dynamics in the presence
of noncanonical kinetic terms of the scalar field, namely ultra slow-roll G-inflation.
We compute the evolution of
the curvature perturbation and show that the primordial power spectrum
follows a broken power law with an oscillation feature.
It is demonstrated that this could explain the lack of large-scale power 
in the cosmic microwave background temperature anisotropies.
We also point out that the violation of the null energy condition is prohibited
in ultra slow-roll G-inflation and hence a blue tensor tilt is impossible
as long as inflation is driven by the potential.
This statement is, however, not true if the energy density is dominated by the kinetic energy
of the scalar field.
\end{abstract}

\preprint{RUP-16-9}
\maketitle
%%%%%%%%%%%%%%%%%%%%%%%%%%%%%%%%%%%%%%%%%%%%%%%%%%%%%%%%%%%%%%%%%%%%%%
\section{Introduction}

Inflation is a very successful scenario of the early universe
that resolves the problems in the standard big bang cosmology~\cite{Guth:1981, Starobinsky:1980, Sato:1981}.
Conventional inflation models employ
a scalar field $\phi$ (the inflaton) rolling slowly on a nearly flat potential, $V(\phi)$.
The energy density $\rho$ and the pressure $p$ of the scalar field are
then given by $\rho\simeq V$ and $p\simeq -V$, leading to a quasi-de Sitter expansion.
Quantum fluctuations in $\phi$
are generated during inflation~\cite{Mukhanov:1981xt},
which nicely provide the seeds for the large-scale structure of the Universe.
Standard single-field slow-roll inflation indeed predicts
nearly scale-invariant, adiabatic, and Gaussian density perturbations
consistent with the observed cosmic microwave background (CMB) anisotropies~\cite{Planck:2013jfk,Planck2},
though the agreement is not perfect and some anomalies have been reported in CMB data such as
the lack of large-scale power~\cite{Planck2}.

Slow-roll inflation driven by a canonical scalar field is
no doubt very attractive, but it is not the only option.
For example, the inflaton sector of the Lagrangian could contain
non-canonical kinetic terms and/or multiple fields. It is also possible that
the inflaton dynamics is away from slow roll.
Those different possibilities yield different predictions for CMB power spectrum features
and thus can be tested against observations.
It would also be interesting if non-standard models of inflation could explain the observed CMB anomalies.

Since the deviation from the exactly scale-invariant power spectrum
is characterized by the slow-roll parameters,
the violation of the slow-roll approximation would imply
a large tilt as well as a non-de Sitter expansion.
This common wisdom, however, is not true.
If the inflaton is subject to the so-called ``ultra slow-roll'' (or ``nonattractor'')
dynamics~\cite{Kinney:2005vj,Inoue:2001zt},
one of the slow-roll parameters is of ${\cal O}(1)$, but the Universe
undergoes a quasi-de Sitter phase.
In the ultra slow-roll phase, the curvature perturbation grows on superhorizon scales,
and this growing mode has a scale-invariant spectrum.
Ultra slow-roll inflation is unique in that
large local
non-Gaussianity is produced due to the superhorizon growth despite a single field
model~\cite{Namjoo:2012aa,Martin:2012pe,Huang:2013lda,Motohashi:2014ppa,Mooij:2015yka}.
The same behavior of the curvature perturbation is also found
in the other backgrounds such as matter bounce~\cite{Finelli:2001sr,Wands:1998yp,Allen:2004vz,Cai:2009fn}
and a variant of Galilean genesis~\cite{Liu:2011ns,Piao:2010bi,Nishi:2015pta}.

In this paper, we explore the consequences of non-canonical kinetic terms on
the ultra slow-roll dynamics of inflation
in order to construct a phenomenological model that can explain the CMB anomalies.
In Refs.~\cite{Chen:2013aj,Chen:2013eea}
ultra slow-roll models have already been generalized to
the Lagrangian containing higher power kinetic terms, $(\partial_\mu\phi\partial^\mu\phi)^n$,
{\em i.e.,} the k-essence Lagrangian~\cite{ArmendarizPicon:1999rj},
with a particular emphasis on the violation of Maldacena's consistency relation
for the bispectrum in the squeezed limit~\cite{Maldacena:2002vr}.
In this paper, we allow the inflaton Lagrangian to depend on second derivatives of the scalar
field~\cite{Deffayet:2010qz,Kobayashi:2010cm}.
That is, we consider {\em ultra slow-roll G-inflation}.
Provided that the duration of inflation is ``just enough,''
the primordial power spectrum is of the broken power-law form
having a blue tilt on large scales in ultra slow-roll k/G-inflation.
We would therefore point out that
this could explain the large-scale suppression of the CMB power.

G-inflation is an interesting class of models that admits in principle
a blue spectrum of primordial tensor modes
by violating the null energy condition stably~\cite{Kobayashi:2010cm}
(see, however, Ref.~\cite{Cai:2014uka}).
Therefore, it should be clarified in which concrete models this is indeed possible.
Under the usual slow-roll conditions, it was shown
in Ref.~\cite{Kamada:2010qe} that stable potential-driven models cannot produce
a blue tensor spectrum. This result leads to the question whether
stable violation of the null energy condition is possible 
under the ultra slow-roll conditions,
which also motivates the study of the combined scenario of ultra slow-roll inflation
and G-inflation.

%{Kamada:2010qe} 

The paper is organized as follows. In the next section
we introduce ultra slow-roll G-inflation and study
the evolution of the curvature perturbation.
The suppressed CMB power due to the broken power-law primordial spectrum
is demonstrated.
We then consider the inflationary universe
approaching a kinetically driven de Sitter attractor
and show that a blue spectrum of primordial tensor modes is possible
in Sec.~III. We draw our conclusions in Sec.~IV.

\section{Ultra slow-roll G-inflation}

\subsection{The background equations}

Let us consider the action of the form
\begin{align}
&S=\int\D^4x\sqrt{-g}
\left[
\frac{\mpl^2}{2}R+{\cal L}_\phi
\right],
\\
&{\cal L}_\phi=-V(\phi)+K(X)-G(X)\Box\phi,\label{Lphi1}
\end{align}
where $K$ and $G$ are arbitrary functions of
$X:=-(1/2)g^{\mu\nu}\partial_\mu\phi\partial_\nu\phi$.
This action was obtained by generalizing
the Galileon~\cite{Nicolis:2008in},
and the first application to dark energy/inflation was done in Refs.~\cite{Deffayet:2010qz,Kobayashi:2010cm}.
A further generalization yields generalized G-inflation~\cite{Kobayashi:2011nu}, {\em i.e.,}
inflation in the Horndeski theory~\cite{Horndeski:1974wa,Deffayet:2011gz,Charmousis:2011bf},
but just for simplicity
we focus on the above subclass; G-inflation captures important aspects of generalized G-inflation
and in the following analysis
extending the former to the latter would be more or less straightforward.

For a flat Friedmann universe,
$\D s^2=-\D t^2+a^2(t)\delta_{ij}\D x^i\D x^j$,
the field equations read
\begin{align}
&3\mpl^2 H^2=V-K+2XF,\label{bg1}
\\
&\mpl^2\dot H=-XF+XG_X\ddot\phi,\label{bg2}
\\
&\frac{\D}{\D t}\left(\dot\phi F\right)+3H\dot\phi F+\frac{\D V}{\D \phi}=0,\label{bg3}
\end{align}
where
\begin{align}
F:=K_X+3G_X H \dot\phi,
\end{align}
and a subscript $X$ denotes differentiation in terms of $X$.
For a canonical model of inflation, we have $K=X$, $G=0$, so that $F=1$,
and in this case Eqs.~(\ref{bg1})--(\ref{bg3}) reproduce the standard equations.

{\em Slow-roll} G-inflation presumes
that the energy density is dominated by the potential in Eq.~(\ref{bg1})
and the friction term balances the slope of the potential in Eq.~(\ref{bg3}),
{\em i.e.},
\begin{align}
3\mpl^2H^2\simeq V,
\quad
3H\dot\phi F+\frac{\D V}{\D \phi}\simeq 0.
\end{align}
This class of G-inflation models has been investigated in~\cite{Kamada:2010qe}
with an emphasis on the consequence of a non-canonical kinetic term with $F\gg 1$.

In this paper, we propose {\em ultra slow-roll} G-inflation
for which the potential is {\em exactly} flat, $V=V_0=$ const.
This is an extension of the notion of ultra slow-roll inflation~\cite{Kinney:2005vj,Inoue:2001zt}
to the models with noncanonical kinetic terms.
Ultra slow-roll k-inflation
has been previously discussed in Refs.~\cite{Huang:2013lda,Chen:2013aj,Chen:2013eea}.
When $V=$ const, the friction term cannot balance the potential slope
because $\D V/\D\phi =0$,
and as a result the scalar field equation of motion yields
\begin{align}
\frac{\D}{\D t}\left(\dot\phi F\right)+3H\dot\phi F=0
\quad\Rightarrow\quad \dot\phi F\propto a^{-3}.\label{beki}
\end{align}
It is still assumed that $V_0\gg |K|,\;|XF|$,
and hence $3\mpl^2H^2\simeq V_0$.
Note that even in the case of an extremely flat potential
ultra slow-roll inflation is possible. However, in the present paper
we focus on the exactly flat potential just for simplicity and clarity.

To be more specific, let us consider
the case with $F\propto X^{p-1}$.
Here, $p\,(\ge 1)$ is not necessarily an integer.
It follows from Eq.~(\ref{beki}) that
\begin{align}
\dot\phi\propto a^{-3/(2p-1)},
\quad
\ddot\phi=-\frac{3H\dot\phi}{2p-1}\propto a^{-3/(2p-1)}.
\end{align}
Then, Eq.~(\ref{bg2}) reduces to
\begin{align}
-\mpl^2\dot H =X\left(F+\frac{3}{2p-1}G_XH\dot\phi\right).
\end{align}
Since $|F|\gtrsim|G_XH\dot\phi|$ (barring from accidental cancellation
of the two terms in $F$),
we have
\begin{align}
\epsilon:=-\frac{\dot H}{H^2}={\cal O}(|XF|/V_0)\ll 1,
\end{align}
and
\begin{align}
\epsilon\propto X^p\propto a^{-6p/(2p-1)}.
\end{align}
The second slow-roll parameter, $\eta:=\dot\epsilon/H\epsilon$,
is given by
\begin{align}
\eta=-\frac{6p}{2p-1}={\rm const}<0,\label{eta-formula}
\end{align}
where $-6\le \eta < -3$ for $p\ge 0$ and hence $|\eta|$ is not small.
For example, the k-inflation model with $K\propto X^{n}$ and $G=0$
gives $p=n$, while the model with $K=0$ and $G\propto X^{n}$ leads to
$p=n+1/2$.
In the latter case $p$ is a half-integer if $n$ is an integer.
Substituting $p=1$ to Eq.~(\ref{eta-formula}),
one can recover the previous result of ultra slow-roll inflation, $\eta=-6$.

%%%%%%%%%%%%%%%%%%%%%%%%%
\begin{figure}[tb]
  \begin{center}
    \includegraphics[keepaspectratio=true,height=95mm]{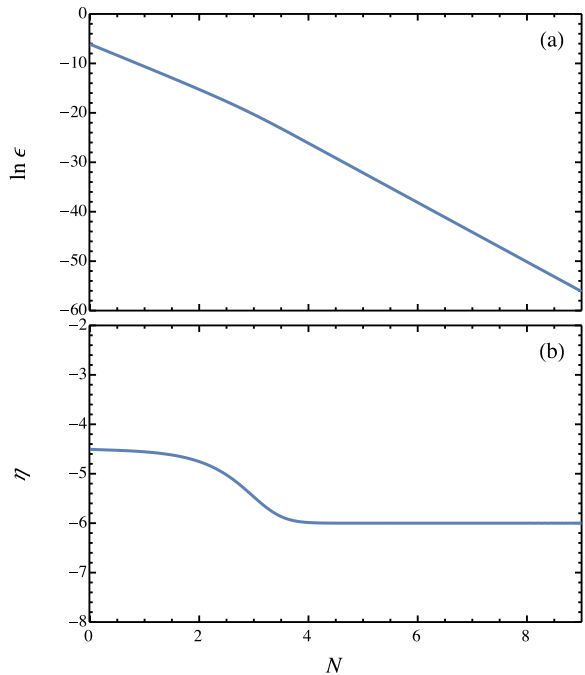}
  \end{center}
  \caption{The evolution of the slow-roll parameters, $\epsilon$ (a) and $\eta$ (b),  in
  the two-phase model of ultra slow-roll G-inflation as functions of
  $e$-folding number $N$.
  }%
  \label{fig:plt1.eps}
\end{figure}
%%%%%%%%%%%%%%%%%%%%%%%%%

Bearing in mind the above ultra slow-roll dynamics with a noncanonical kinetic term,
let us suppose that
${\cal L}_\phi$ is composed of the standard kinetic term plus
some higher-power term, as
\begin{align}\label{k-type}
{\cal L}_\phi=-V_0+X+\frac{X^n}{M^{4(n-1)}}\quad (n\ge2),
\end{align}
or
\begin{align}
{\cal L}_\phi=-V_0+X\mp \frac{X^n}{M^{4n-1}}\Box\phi\quad(n\ge1),
\end{align}
and in the early stage of inflation
the higher-power term is much larger than $X$ (but smaller than $V_0$).
During this early stage, the Universe undergoes
ultra slow-roll G-inflation (or ultra slow-roll k-inflation) with $\eta\neq -6$
and $\dot\phi$ decreases as $\dot\phi\propto a^{3+\eta}$.
The higher-power term immediately becomes smaller than the standard kinetic energy, $X$, and
then the Universe switches to the conventional ultra slow-roll phase with $\eta=-6$.
It is thus easy to incorporate the automatic transition between the two phases
of inflation in this setup.
In this paper, we study such two-phase models of
ultra slow-roll inflation.
We dub the early phase
as USRK1$_\eta$ if it is
governed by $K\sim X^n$ and
as USRG1$_\eta$ if governed by $G\sim X^n$,
while we call the second ultra slow-roll phase simply USR2.
A numerical example of the background evolution for
\begin{align}
\label{g-type}
{\cal L}_\phi=-V_0+X-\frac{X}{M^3}\Box\phi,
\end{align}
{\em i.e.,} the transition from the USRG1$_{-9/2}$ phase to the USR2 phase,
is shown in Fig.~\ref{fig:plt1.eps}.

\subsection{Cosmological perturbations}

We assume that the duration of the second phase is just enough,
so that the first phase can be probed by the cosmological perturbations on the largest scales
in the CMB observations.
It has been known that in the previous models of ultra slow-roll inflation
the curvature perturbation grows, contrary to folklore, on superhorizon scales.
As we will see below, this also happens in the case of ultra slow-roll G-inflation,
though in the two-phase model we have to be more careful
about the evolution of the curvature perturbation across the transition
between the two phases.

The quadratic action for the curvature perturbation $\zeta$
on uniform $\phi$ hypersurfaces is given by~\cite{Deffayet:2010qz,Kobayashi:2010cm,Kobayashi:2011nu}
\begin{align}
S^{(2)}_\zeta=\int \D t\D^3x a^3\left[{\cal G}_S\dot\zeta^2-\frac{{\cal F}_S}{a^2}(\partial\zeta)^2\right],
\label{qaczeta}
\end{align}
where
\begin{align}
{\cal F}_S&=\frac{\mpl^4}{a}\frac{\D}{\D t}\left(\frac{a}{\Theta}\right)-\mpl^2,
\\
{\cal G}_S&=\frac{\mpl^4\Sigma }{\Theta^2}+3\mpl^2,
\end{align}
with
\begin{align}
\Sigma &:=XK_X+2X^2K_{XX}+12H\dot\phi XG_X
\notag \\ &\quad
+6H\dot\phi X^2G_{XX}-3\mpl^2H^2,
\\
\Theta&:=\mpl^2H-\dot\phi XG_X.
\end{align}
By using these expressions, the sound speed of the curvature perturbation can be written as
$c_s=\sqrt{{\cal F}_S/{\cal G}_S}$.

Let us first consider the USRK1$_\eta$ phase.
In this phase, we have
\begin{align}
{\cal F}_S=\mpl^2\epsilon,
\quad
{\cal G}_S=\frac{\mpl^2\epsilon}{-1-\eta/3},\label{usrk1-fg}
\end{align}
so that $c_s^2=-1-\eta/3$.
Since $\eta<-3$, the ghost and gradient instabilities
can be avoided if $\epsilon >0$.
In the USRG1$_\eta$ phase, we find
\begin{align}
{\cal F}_S=\frac{\mpl^2\epsilon}{-\eta},
\quad
{\cal G}_S=\frac{9\mpl^2\epsilon}{\eta(3+\eta)},\label{usrg1-fg}
\end{align}
and therefore $c_s^2=(-3-\eta)/9$.
The ghost and gradient instabilities can thus be avoided
provided that $\epsilon > 0$.
Finally, the result in the USR2 phase can be obtained by
substituting $\eta=-6$ to Eq.~(\ref{usrk1-fg}),
\begin{align}
{\cal F}_S={\cal G}_S=\mpl^2\epsilon.\label{usr2-fg}
\end{align}

When the galileon-like $G\Box\phi$ term comes into play,
the stability conditions at the level of linear perturbations
are in general uncorrelated with
the sign of $\epsilon$ or, equivalently, the sign of $\dot H$~\cite{Deffayet:2010qz,Kobayashi:2010cm,Kobayashi:2011nu}.
Therefore, the null energy condition can in principle be violated stably.
However, in the potential-driven inflation models it has been shown,
{\em under the slow-roll approximation},
that the stability conditions require $\epsilon>0$~\cite{Kamada:2010qe}.
Here, we have generalized the previous statement and
shown that the stability of the potential-driven models
amounts to $\epsilon>0$ even in the case of {\em ultra slow-roll},
where the different approximation is made.
Note that nonlinear stability is yet unclear~\cite{Sawicki:2012pz}.

In all three cases described by Eqs.~(\ref{usrk1-fg})--(\ref{usr2-fg}),
the functions ${\cal F}_S$ and ${\cal G}_S$ are of the form
\begin{align}
{\cal F}_S=\mpl^2 f(\eta)\epsilon,
\quad
{\cal G}_S=\mpl^2 g(\eta)\epsilon.
\end{align}
Let us look at the solutions to the equation of motion for $\zeta$
in each $\eta=\;$const phase of
the ultra slow-roll background, $\epsilon \propto a^{\eta}$.
The equation of motion in the Fourier space reduces to
\begin{align}
\zeta_k''-\frac{2+\eta}{\tau}\zeta_k'+c_s^2k^2\zeta_k=0,\label{perteq}
\end{align}
where a dash stands for differentiation with respect to
the conformal time $\tau=-1/aH$ and $c_s^2=f/g=\;$const.
The linearly independent solutions are given by
$\psi_k$ and its complex conjugate, $\psi_k^\ast$, where
\begin{align}
\psi_k&=\frac{1}{2}\sqrt{\frac{\pi}{2}}\frac{H}{\mpl k^{3/2}}
\frac{g^{1/4}}{f^{3/4}\epsilon^{1/2}}
\notag \\&\quad\times
(-c_sk\tau)^{3/2}
H_{(3+\eta)/2}^{(1)}(-c_sk\tau),
\end{align}
and $H_\nu^{(1)}$ is the Hankel function of the first kind.
Note the normalization of $\psi_k$,
\begin{align}
\psi_k\approx
\frac{1}{a\left(2{\cal G}_S\right)^{1/2}}\cdot \frac{e^{-ic_sk\tau}}{\sqrt{2c_sk}}
\quad (|c_sk\tau|\to\infty),\label{advac}
\end{align}
up to a phase factor.
In particular, in the USR2 phase
the linearly independent solutions are given by $y_k$ and $y_k^\ast$, where
\begin{align}
y_k=\frac{1}{2}\sqrt{\frac{\pi}{2}}\frac{H}{\mpl k^{3/2}\epsilon^{1/2}}(-k\tau)^{3/2}H_{-3/2}^{(1)}(-k\tau).
\end{align}

On superhorizon scales ($|c_sk\tau|\ll 1$), we find, for $3+\eta<0$,
\begin{align}
\psi_k\simeq
\frac{A}{2}\frac{H}{\mpl\epsilon^{1/2}}\frac{g^{1/4}}{f^{3/4}}
|c_s\tau|^{3+\eta/2}k^{(3+\eta)/2},\label{assympsi}
\end{align}
where
\begin{align}
A:=2^{-\nu}\sqrt{\frac{\pi}{2}}\left[
\frac{1}{\Gamma(1+\nu)}+\frac{\cos(\pi\nu)\Gamma(-\nu)}{\pi i}\right],
\end{align}
with
\begin{align}
\nu:=\frac{3+\eta}{2}.
\end{align}
From Eq.~(\ref{assympsi}) we see that
\begin{align}
\psi_k\propto \frac{|\tau|^{3+\eta/2}}{\epsilon^{1/2}}\propto |\tau|^{3+\eta},
\end{align}
and hence the curvature perturbation grows on superhorizon scales.

In the two-phase model of ultra slow-roll inflation, Eq.~(\ref{advac}) shows that
the appropriate initial condition deep in the USRK1$_\eta$/USRG1$_\eta$ phase
is given by
\begin{align}
\zeta_k\to \psi_k.
\end{align}
One can then solve Eq.~(\ref{perteq}) numerically to
obtain the power spectrum of $\zeta_k$ at the end of inflation.
Here we provide a complementary argument
for evaluating analytically the late-time amplitude of $\zeta_k$.
To do so, let us assume that the transition between the two phases
occurs instantaneously at some $\tau=\tau_\ast$.
The solution in the USR2 phase ($\tau>\tau_\ast$) can be written as
\begin{align}
\zeta_k=\alpha_ky_k+\beta_ky_k^\ast,
\end{align}
where the coefficients $\alpha_k$ and $\beta_k$
are fixed by matching the curvature perturbation at $\tau=\tau_\ast$.
Noting that ${\cal G}_S=\mpl^2g(\eta)\epsilon$ and $\epsilon$
is continuous while $g(\eta)$
undergoes a sudden change $g\to 1$
across the transition,
the matching conditions are summarized as~\cite{Deruelle:1995kd,Nishi:2014bsa}
\begin{align}
\zeta_k|_{\tau_\ast-0}=\zeta_k|_{\tau_\ast+0},
\quad
g(\eta)\zeta_k'|_{\tau_\ast-0}
=\zeta_k'|_{\tau_\ast+0},
\end{align}
where $g=1/(-1-\eta/3)$ for USRK1$_\eta$ and
$g=9/\eta(3+\eta)$ for USRG1$_\eta$, as we have mentioned.
We thus obtain
\begin{align}
\alpha_k&=\frac{\pi i}{4}\frac{|k\tau_\ast|}{g^{1/2}}\bigl[
H_\nu^{(1)}(-c_sk\tau_\ast)H_{-5/2}^{(2)}(-k\tau_\ast)
\notag \\ &\qquad\qquad\quad
-c_sgH_{\nu-1}^{(1)}(-c_sk\tau_\ast)H_{-3/2}^{(2)}(-k\tau_\ast)
\bigr],\label{calpha}
\\
\beta_k&=-\frac{\pi i}{4}\frac{|k\tau_\ast|}{g^{1/2}}\bigl[
H_\nu^{(1)}(-c_sk\tau_\ast)H_{-5/2}^{(1)}(-k\tau_\ast)
\notag \\ &\qquad\qquad\quad
-c_sgH_{\nu-1}^{(1)}(-c_sk\tau_\ast)H_{-3/2}^{(1)}(-k\tau_\ast)
\bigr],\label{cbeta}
\end{align}
and using those coefficients the power spectrum is obtained as
\begin{align}
{\cal P}_\zeta(k)=\frac{|\alpha_k+\beta_k|^2}{8\pi^2}\frac{H^2}{\mpl^2\epsilon}.\label{psan}
\end{align}

In the $|k\tau_\ast|\ll 1$ limit, we find
\begin{align}
|\alpha_k+\beta_k|^2\simeq
\frac{2^{-2-\eta}\Gamma^2(1-\nu)}{9\pi}c_s^{3+\eta}g|k\tau_\ast|^{6+\eta}.\label{amp1}
\end{align}
This corresponds to the modes that exit the horizon
during the USRK1$_\eta$/USRG1$_\eta$ phase.
Equation~(\ref{amp1}) shows that for $|k\tau_\ast|\ll 1$
\begin{align}
n_s-1=6+\eta,
\end{align}
and thus the power spectrum is blue on large scales.

In the opposite limit, $|k\tau_\ast|\gg 1$, we have
\begin{align}\label{amp2}
|\alpha_k+\beta_k|^2\approx
c_s g+\frac{1-c_s^2 g^2}{c_s g}\sin^2(k\tau_\ast).
\end{align}
It can be seen that the power spectrum oscillates even at large $k$~\cite{Nakashima:2010sa}.
However, the oscillation is due to the artifact of the
sudden change approximation enforcing infinitely large $\dot \eta/H\eta$ and $\dot c_s/Hc_s$ at the transition,
which leads to the nonadiabatic evolution
of all the modes.\footnote{We further discuss the relation between the sudden change approximation
and the oscillating spectrum at large $k$ in Appendix~\ref{append}.}
From the numerical example in Fig.~\ref{fig:plt1.eps}
we see that the actual transition occurs smoothly
on a time scale of $N={\cal O}( 1)$.
In this case, the modes with sufficiently large $k$
remain in their adiabatic vacuum, yielding
\begin{align}
|\alpha_k+\beta_k|^2\approx 1.
\end{align}

\begin{figure}[tb]
  \begin{center}
    \includegraphics[keepaspectratio=true,height=50mm]{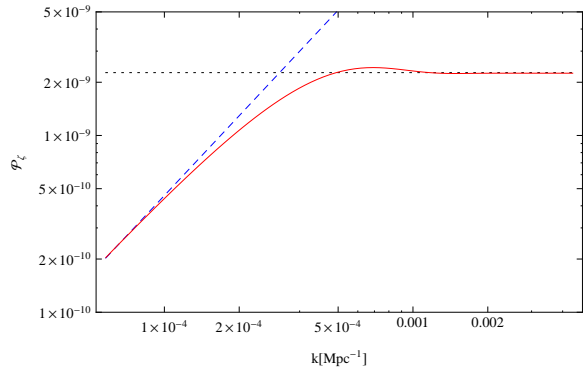}
  \end{center}
  \caption{The power spectrum of the primordial curvature perturbations 
  for the model given by Eq. (\ref{g-type}).
  The amplitude and the scales are taken so as to be consistent with the CMB observation.
  }%
  \label{fig:USRG_power.eps}
\end{figure}

\begin{figure}[tb]
  \begin{center}
    \includegraphics[keepaspectratio=true,height=50mm]{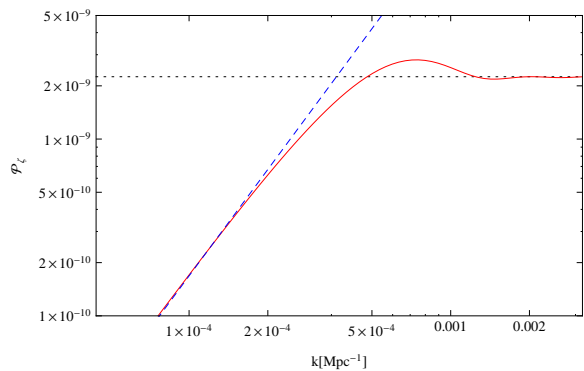}
  \end{center}
  \caption{The power spectrum of the primordial curvature perturbations 
  for the model given by Eq. (\ref{k-type}) with $n = 2$.
  The amplitude and the scales are taken so as to be consistent with the CMB observation.
  }%
  \label{fig:USRK_power.eps}
\end{figure}

Full numerical solutions confirm the above argument.
As examples, we consider two simple models where the Lagrangians are respectively given by Eq.~(\ref{g-type}), that is, USRK1$_{-9/2}$, and Eq. (\ref{k-type}) with $n = 2$, that is, USRG1$_{-4}$.
We plot the power spectrum of the primordial curvature perturbation computed numerically
  for each model (Fig.~\ref{fig:USRG_power.eps} for the two-phase model composed of USRG1$_{-9/2}$
  and USR2, and Fig.~\ref{fig:USRK_power.eps} for that composed of USRK1$_{-4}$
  and USR2). 
  In both figures, the red lines represent the numerical results, while
  the blue dashed lines correspond to the analytic results for $|k \tau_*| \ll 1$
   under the sudden transition approximation.
  The black dotted lines are given by ${\mathcal P}_\zeta \simeq H^2 / (8 \pi^2 M_{\rm Pl}^2 \epsilon)$,
  {\em i.e.}, $|\alpha_k+\beta_k|\approx 1$.
   The amplitude and the scales are taken so as to be consistent with the CMB observation,
  by choosing the inflationary Hubble scale and the scale $M$ appropriately.
From these figures, 
we find that suppression of the power on large scales can be realized
due to the fact that the USRG1$_{-9/2}$/USRK1$_{-4}$ phase exists prior to the USR2 phase
which produces the scale-invariant curvature perturbation.
Furthermore, we can see that our analytic formula~(\ref{psan}) with
the coefficients~(\ref{calpha}) and~(\ref{cbeta})
reproduces the numerical result
except that
the oscillating feature for large $k$ appears in Eq.~(\ref{amp2}).
This is because 
the actual transition occurs smoothly
on a time scale of $N={\cal O}( 1)$, as we have discussed.
  Note that since $\dot c_s/Hc_s, \dot\eta/H\eta\sim 1$ at the transition,
   a few oscillations are still found around the break in the spectrum.

In the present setup, the equation of motion for
the tensor perturbations remains the same as the standard one.
Therefore, the power spectrum of the primordial tensor modes is given by
\begin{align}
{\cal P}_h=\frac{8}{\mpl^2}\left.\left(\frac{H}{2\pi}\right)^2\right|_{k=aH},
\quad
n_t=-2\epsilon.\label{tensorsp}
\end{align}
Since $\epsilon>0$ is required for stability, the tensor power spectrum is
always red tilted.

Before closing this subsection,
it is appropriate to give a short comment
on potential drawbacks that have not been discussed so far.
First, the present model does not have a mechanism to end the period of inflation.
Second, the curvature perturbation generated in the second phase has
an exactly scale-invariant spectrum, while observations imply that $n_s\simeq 0.96$.
Both of the drawbacks stem from the exactly flat potential.
Introducing a mild slope of the potential~\cite{Chen:2013aj,Chen:2013eea},
this issue is expected to be evaded.

%%%%%%%%%%%%%%%%%%%%%%%%%
%%%%%%%%%%%%%%%%%%%%%%%%% 

%%%%%%%%%%%%%%%%%%%%%%%%%
%%%%%%%%%%%%%%%%%%%%%%%%% 

\subsection{Suppression of CMB power on the largest scales}

%%%%%%%%%%%%%%%%%%%%%%%%%
\begin{figure}[htbp]
  \begin{center}
    \includegraphics[keepaspectratio=true,height=60mm]{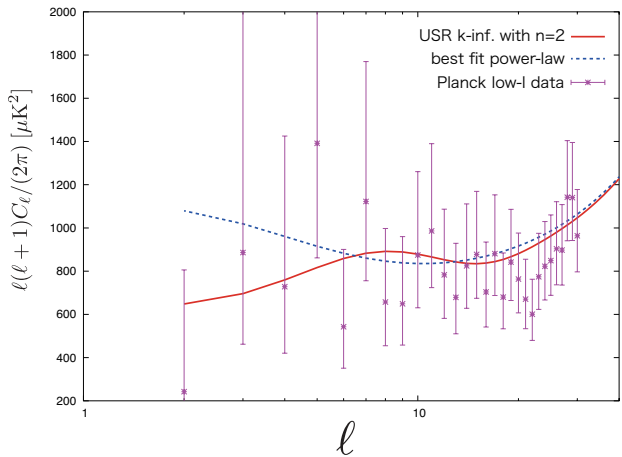}
  \end{center}
  \caption{CMB angular power spectrum calculated from the primordial power spectrum shown in
  Fig. \ref{fig:USRK_power.eps} (red solid line) with Planck low-$\ell$ data~\cite{Planck2} (purple dots with error bars)
  and the power spectrum for the best fit power-law $\Lambda$CDM model (blue dotted line).
  }%
  \label{fig:cmb_temp.eps}
\end{figure}
%%%%%%%%%%%%%%%%%%%%%%%%% 

Given that the power spectrum of the primordial curvature perturbation
has the broken power-law form,
let us briefly discuss the CMB temperature power spectrum in ultra slow-roll G-inflation.

Our model relies on the assumption that
there are two phases during inflation and
the duration of the second phase
is just enough. A number of attempts in this direction have been done so far
to explain the lack of large-scale power in the
CMB, e.g., by changing the inflaton potential or considering the pre-inflationary era  
~\cite{Contaldi:2003zv,Kawasaki:2003dd,Wang:2007ws,Scardigli:2010gm,BouhmadiLopez:2012by,
Kouwn:2014aia,Jain:2008dw,Dudas:2012vv,Namjoo:2012xs,White:2014aua,Biswas:2013dry,Chen:2015gla}.
Comparing with these possibilities, it should be emphasized that in our model the first phase too is quasi-de Sitter.
Nevertheless, the primordial spectrum is tilted on the largest scales.

As an example, Fig.~\ref{fig:cmb_temp.eps} shows the CMB angular power spectrum calculated from the
primordial power spectrum shown in Fig.~\ref{fig:USRK_power.eps} (red solid line).
In this figure, we also plot the Planck low-$\ell$ data~\cite{Planck2} (purple dots with error bars)
and  the power spectrum for the best fit power-law $\Lambda$CDM model (blue dotted line).  
The angular power spectrum is calculated by using the CLASS code~\cite{Blas:2011rf}.
 From this figure, we find that the suppression of CMB power on the largest scales
 can indeed be realized in our model.
Using $H_{\rm inf} = 2.7 \times 10^{-39} \mpl$ and $M = 6.8 \times 10^{-23} \mpl$, we find that 
this model roughly improves the effective $\chi^2$ by $\Delta \chi^2_{\mathrm{eff}} \approx -3.4$.

\section{Approaching de Sitter Universe dominated by kinetic energy}
\label{sec:bluetensor}

So far we have studied ultra slow-roll inflation driven by
the constant potential energy of a scalar field. Its kinetic energy is much
smaller than the potential energy and dilutes as the Universe expands
with a nearly constant Hubble rate.
Here one would notice that
quasi-de Sitter inflation can be driven as well by nearly constant kinetic energy
of a noncanonical scalar field.
In this section, we study inflationary expansion
that is similar to ultra slow-roll inflation in the sense that
the Universe approaches the de Sitter attractor, but in the present case
the Universe is
dominated by the kinetic energy rather than the potential.

The scalar-field Lagrangian we consider is
\begin{align}
{\cal L}_\phi=K(X)-G(X)\Box\phi,
\end{align}
where the potential term is removed from Eq.~(\ref{Lphi1}).
The background equations are thus given by
\begin{align}
3\mpl^2H^2&=-K+\dot\phi J,
\\
\mpl^2\dot H&=-\frac{1}{2}\dot\phi J+X G_X\ddot\phi,
\\
\dot J+3H J&=0,
\end{align}
where
we write
\begin{align}
J:=\dot\phi F=\dot\phi K_X+6XG_XH.
\end{align}

We introduce a small parameter $\xi$ and assume the following ansatz:
\begin{align}
H=H_0+\delta H(t),
\quad
\dot\phi=\dot\phi_0+\dot{\delta\phi}(t),
\end{align}
where $H_0$ and $\dot\phi_0$ are constants and
$\{\delta H,\dot{\delta\phi}\}={\cal O}(\xi)\times\{H_0,\dot\phi_0\}$.
At leading order,
\begin{align}
3\mpl^2H^2=-K,
\quad
J=0.
\end{align}
Those two equations algebraically determine $H_0$ and $\dot\phi_0$,
yielding de Sitter inflation, $a\propto e^{H_0 t}$.
Then, at ${\cal O}(\xi)$ we have
\begin{align}
6\mpl^2H_0\delta H&=-\dot\phi_0K_X\dot{\delta\phi}+\dot\phi_0\delta J,\label{bgeq21}
\\
\mpl^2\dot{\delta H}&=-\frac{1}{2}\dot\phi_0\delta J+X_0G_X\ddot{\delta\phi},\label{bgeq22}
\\
\dot{\delta J}+3H_0 \delta J&=0,\label{bgeq23}
\end{align}
where $X_0:=\dot\phi_0^2/2$,
\begin{align}
\delta J &=\left[K_{X}+2XK_{XX}+6H\dot\phi\left(G_{X}+XG_{XX}\right)\right]\dot{\delta \phi}
\notag \\ &\quad
+\left[6XG_{X}\right]\delta H,\label{def-deltaJ}
\end{align}
and the quantities in the square brackets in Eq.~(\ref{def-deltaJ})
are evaluated at $H=H_0$ and $\dot\phi=\dot\phi_0$.
Note that Eq.~(\ref{bgeq22}) [Eq.~(\ref{bgeq23})]
can be derived from Eq.~(\ref{bgeq23}) [Eq.~(\ref{bgeq22})] and Eq.~(\ref{bgeq21}).

We have consistent solutions to Eqs.~(\ref{bgeq21})--(\ref{bgeq23})
satisfying
\begin{align}
&\delta H\propto a^{-3},
&&\dot{\delta\phi}\propto a^{-3},
\\
&\dot{\delta H}=-3H_0\delta H \propto a^{-3},
&&\ddot{\delta\phi}=-3H_0\dot{\delta\phi}\propto a^{-3},
\end{align}
characterized by one integration constant.
The slow-roll parameter is given by
\begin{align}
\epsilon=-\frac{\dot H}{H^2}\simeq -\frac{\dot{\delta H}}{H_0^2}
=\frac{3\delta H}{H_0}\propto a^{-3},
\end{align}
and hence $\epsilon$ is of ${\cal O}(\xi)$.
However, the second slow-roll parameter is not small: $\eta=-3$.
Introducing a constant ${\cal C}$ [$={\cal O}(\xi)$],
we write $\epsilon={\cal C}/a^3$, and then $\delta H$ and $\dot{\delta \phi}$
can be written in terms of ${\cal C}$ as
\begin{align}
\delta H=\frac{H_0}{3}\frac{{\cal C}}{a^3},
\end{align}
and
\begin{align}
\dot{\delta\phi}=
\frac{H_0\Theta_0}{X_0\Gamma_0}\frac{{\cal C}}{a^3}.
\end{align}
where
\begin{align}
\Theta_0&:=\mpl^2H_0-\dot\phi_0X_0G_X,
\\
\Gamma_0&:=\dot\phi_0K_{XX}+6H_0(G_X+X_0G_{XX}).
\end{align}
Note that,
at this stage,
there is no privileged choice of the sign of ${\cal C}$
, and hence $\epsilon$ can be negative,
because ${\cal C}$ is just an integration constant.

Let us move to discuss the behavior of cosmological perturbations
around this quasi-de Sitter background.
Also in this case the tensor perturbations obey
the standard formula, so that the primordial power spectrum is given by
Eq.~(\ref{tensorsp}).
Therefore, the tensor amplitude is determined at leading order
as ${\cal P}_h\simeq 2H_0^2/\pi^2\mpl^2$, and
its tilt is given in terms of the ${\cal O}(\xi)$ quantity as
$n_t=-2\epsilon$.

The quadratic action for the curvature perturbation is of the form~(\ref{qaczeta})
with
\begin{align}
{\cal F}_S&={\cal F}_{S0}+f_S\frac{{\cal C}}{a^3},
\\
{\cal G}_S&={\cal G}_{S0}+g_S\frac{{\cal C}}{a^3}.
\end{align}
Here, ${\cal F}_{S0}$ and ${\cal G}_{S0}$ are leading-order quantities
and are given explicitly by
\begin{align}
{\cal F}_{S0}&=\frac{\mpl^2\dot\phi_0 X_0G_X}{\Theta_0},
\\
{\cal G}_{S0}&=\frac{\mpl^4X_0}{\Theta^2}\left(K_X+\dot\phi_0\Gamma_0+\frac{6X_0^2G_X^2}{\mpl^2}\right).
\end{align}
The coefficients of the ${\cal O}(\xi)$ terms are also constant
and
\begin{align}
f_S&=\frac{\mpl^4H_0\dot\phi_0}{3\Theta^2_0}\left[
2X_0G_X+\frac{3\Theta_0(K_{XX}+H_0\dot\phi_0 G_{XX})}{\Gamma_0}
\right],
\end{align}
while the form of $g_S$ is messy and involves $G_{XXX}$ as well as $K_{XXX}$.
The stability conditions are given at leading order by
\begin{align}
{\cal F}_{S0}>0,\quad{\cal G}_{S0}>0.
\end{align}
It should be emphasized here that the sign of ${\cal C}$
is not important for stability, as its contribution is subleading.
Therefore, the tensor power spectrum can be blue
on a healthy background, depending on the initial conditions.

The primordial power spectrum reduces to
\begin{align}
{\cal P}_\zeta=\left.\frac{{\cal G}_S^{1/2}}{{\cal F}_S^{3/2}}\frac{H^2}{4\pi^2}\right|_{k=aH/c_s},
\end{align}
and therefore the amplitude is determined by the leading order terms.
The spectral index is
\begin{align}
n_s-1=\left(-2
-\frac{3g_S}{2{\cal G}_{S0}}+\frac{9f_S}{2{\cal F}_{S0}}
\right)\epsilon.
\end{align}
It can be seen that we have a nearly scale-invariant spectrum
even though the second slow-roll parameter is as large as $-3$.
While the stability conditions are given by
the first and second derivatives of $K(X)$ and $G(X)$,
$n_s$ depends on the third derivatives of those functions through $g_S$.
Therefore, using the functional degrees of freedom
one can realize $n_s\simeq 0.96$ on a stable background.

\section{Summary}

Under the motivation of constructing
a phenomenological model to explain the CMB anomalies, we have studied ultra slow-roll models of
G-inflation with a constant potential.
We have considered an earlier phase of inflation where
a higher-power term in $X=-(\partial_\mu\phi\partial^\mu\phi)/2$
governs the scalar-field dynamics
prior to the usual ultra slow-roll phase.
The primordial power spectrum of the curvature perturbation
in such a two-phase model
has been evaluated analytically and numerically to show that
the growing curvature perturbation has a broken power-law spectrum.
The transition between the two phases occurs on a time scale of $N={\cal O}(1)$, and
a few oscillations appear around the break of the spectrum due to
the change of the sound speed $c_s$.
As a consequence of such a feature, 
our model could explain the suppression of CMB power on the largest scales
which has been reported in the CMB observations \cite{Planck2}.

In the presence of the Galileon-like interaction $G(\phi, X)\Box\phi$,
the null energy condition can in principle be violated stably,
which opens up the interesting possibility of a blue primordial tensor spectrum.
However, in Ref.~\cite{Kamada:2010qe} it was shown under the slow-roll approximation
that this is not possible
in general potential-driven inflation models. Motivated by this fact,
in this paper we have extended the previous argument~\cite{Kamada:2010qe} 
and showed that also in ultra slow-roll models the stability conditions
require the null energy condition.
To realize the stable violation of the null energy condition,
we have to give up potential-driven inflation and
consider kinetically driven models.
We have studied the inflationary Universe approaching the
kinetically driven de Sitter attractor in the way similar to potential-driven ultra slow-roll,
and found that in this case the null energy condition can be violated stably.

%--- Acknowledgements ---%--- Acknowledgements ---%--- Acknowledgements ---%
\acknowledgments 
This work was supported in part by
YITP-W-15-16 in workshop JGRG25 and the JSPS, Grant-in-Aid for Scientific
Research No.~24740161 (T.K.), No. 15K17659(S.Y.), and No. 15H05888 (T.K. and S.Y.).
We thank Kazuhiro Yamamoto and Kiyotomo Ichiki for useful discussions.
%--- Acknowledgements ---%--- Acknowledgements ---%--- Acknowledgements ---%

\appendix
%-------------------------------------------------------------------%
\section{The sudden change approximation and the oscillating power spectrum}
\label{append}

%%%%%%%%%%%%%%%%%%%%%%%%%
\begin{figure}[tb]
  \begin{center}
    \includegraphics[keepaspectratio=true,height=45mm]{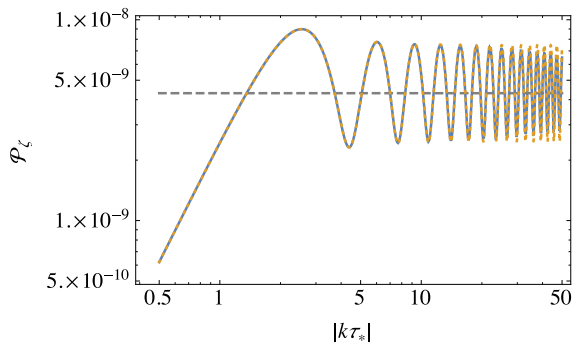}
  \end{center}
  \caption{The power spectrum of $\zeta$ for a sudden transition with $\lambda=100$ (solid curve).
  For comparison the analytic estimate is also shown as a dotted curve.
  For sufficiently large $k$ the spectrum is given by ${\cal P}_\zeta\simeq H^2/(8\pi^2\mpl^2\epsilon)$
  (dashed line), though such behavior is not shown explicitly in the figure.
  }%
  \label{fig:TML100.eps}
\end{figure}
\begin{figure}[tb]
  \begin{center}
    \includegraphics[keepaspectratio=true,height=45mm]{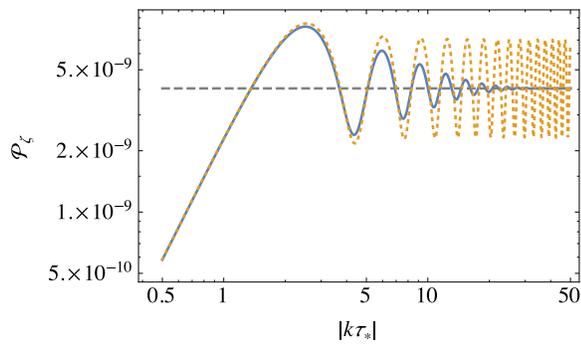}
  \end{center}
  \caption{The power spectrum of $\zeta$ for a mild transition with $\lambda=10$ (solid curve).
  For comparison the analytic estimate is also shown as a dotted curve.
  For sufficiently large $k$ the spectrum is given by ${\cal P}_\zeta\simeq H^2/(8\pi^2\mpl^2\epsilon)$
  (dashed line).
  }%
  \label{fig:TML10.eps}
\end{figure}
%%%%%%%%%%%%%%%%%%%%%%%%%

To see how the sudden change approximation affects the power spectrum of $\zeta$
at large $k$, let us study the toy model of two-phase inflation in which
the ``sharpness'' of the transition is controllable.
The model is characterized by the slow-roll parameter
\begin{align}
\epsilon=\epsilon_\ast
\left\{\cosh[\lambda(N-N_\ast)]\right\}^{(\eta_1-\eta_0)/2\lambda}e^{(\eta_1+\eta_0)(N-N_\ast)/2},
\end{align}
which yields
\begin{align}
\eta = \frac{1}{2}\left\{
(\eta_1-\eta_0)\tanh[\lambda(N-N_0)]+\eta_1+\eta_0
\right\},
\end{align}
with $\epsilon_\ast \ll 1$.
This describes a change from $\eta_0$ to $\eta_1$ at around $N=N_\ast$ on
a quasi-de Sitter background,
and thus mimics the background evolution of the two-phase model of
ultra slow-roll inflation presented in Fig.~\ref{fig:plt1.eps}.
However, in this toy model the ``sharpness'' of the transition can be controlled by
the dimensionless parameter $\lambda$.

The above background can be realized by a suitable choice of the ultra slow-roll k-inflation Lagrangian.
The evolution of the curvature perturbation follows from the quadratic action
with ${\cal F}_S=\mpl^2\epsilon$ and ${\cal G}_S=\mpl^2/(-1-\eta/3)$.
Numerical solutions for
the power spectrum of $\zeta$ evaluated at some later time
are shown in Figs.~\ref{fig:TML100.eps} and ~\ref{fig:TML10.eps}
in the case of the transition from $\eta_0=-4$ to $\eta_1=-6$. The solid curve in
Fig.~\ref{fig:TML100.eps} corresponds to a sudden transition with $\lambda = 100$,
while that in Fig.~\ref{fig:TML10.eps} corresponds to a milder transition with $\lambda = 10$,
showing that in the case of the sudden transition
the oscillation in the power spectrum persists down to small scales.
For a mild transition the analytic formula obtained in the main text
cannot be used on small scales, but instead one may have
${\cal P}_\zeta\simeq H^2/(8\pi^2\mpl^2\epsilon)$.

%-------------------------------------------------------------------%

%---------   References   ---------%

\end{document}